\documentclass[graybox]{svmult}

\usepackage{mathptmx}       
\usepackage{helvet}         
\usepackage{courier}        
\usepackage{type1cm}        
\usepackage{float}
\usepackage{makeidx}         
\usepackage{graphicx}        
\usepackage{multicol}        
\usepackage[bottom]{footmisc}

\makeindex             

\begin{document}

\title*{Semi-convection: What is the underlying physical context~?}

\author{A. Noels}

\institute{A. Noels \at Institut d'Astrophysique et de G\'eophysique, University of Li\`ege, All\'ee du 6 Ao\^ut, 17, Li\`ege, Belgium, \email{Arlette.Noels@ulg.ac.be}}

\maketitle

\abstract{Stellar conditions leading to a possible semi-convective mixing are discussed in three relevant cases: (1) low mass MS stars in which the CNO cycle takes progressively the lead over the PP chain due to the increase in temperature as core hydrogen burning proceeds, (2) massive MS stars which experience a large contribution of the radiation pressure to the total pressure and (3) core helium burning stars for which the production of carbon in the core increases the opacity. A short discussion of semi-convection in terms of instability of non radial modes follows.}

\section{Introduction}\label{intro}

In main sequence stars massive enough to burn hydrogen through the CNO cycle, a convective core is already present at the ZAMS. In most cases, its mass extension is maximum at the ZAMS and then it shrinks with time due to the decrease in opacity resulting from the transformation of hydrogen into helium which rather drastically reduces the density of free electrons. In such a case, the opacity is larger outside the convective core in layers which are richer in hydrogen but as a result of the progressive decrease in mass extension of the convective core; there is neither chemical nor opacity discontinuity at the convective border. In some cases however the convective core mass tends to grow as the star evolves and this leads to the formation of a hydrogen discontinuity at the border, the outer border being hydrogen richer and thus more opaque than the inner border. The layers affected by this, although outside the convective core, are unstable towards convection if the Schwarzschild criterion \cite{schw58} is used to define the convective neutrality, i.e.
\begin{equation}
\nabla_{rad} = (\frac{dlnT}{dlnP})_{rad} >  \nabla_{ad} = \frac{\Gamma_2 - 1}{\Gamma_2}
\end{equation}
where the $\nabla_{rad}$ and $\nabla_{ad}$ refer to radiative and adiabatic temperature gradients and $\Gamma_2$ is the secong adiabatic coefficient, while they are stable when applying the Ledoux criterion \cite{led47} since
\begin{equation}
\nabla_{rad} <  \nabla_{ad} + \frac{\beta} {4-3\beta} \frac{dln\mu}{dlnP}
\end{equation}
where $\beta$ is the ratio of the gas pressure to the total pressure and $\mu$  is the mean molecular weight.
In a pioneering analysis, Schwarzschild and H\"{a}rm \cite{sh58} solved this problem by adding a partial chemical mixing in the so-called semi-convective layers in order to ensure their convective neutrality. The helium abundance in these layers was enriched to just the precise amount required to satisfy the Schwarzschild criterion. A similar partial mixing was adopted by Sakashita and Hayashi \cite{sh61} but with the Ledoux criterion instead.

After recalling a few basic points of stellar structure in section \ref{basic}, I shall discuss the physical conditions leading to such an increase with time of the convective core mass, first in low mass stars (sections \ref{lowmass}) and then in massive stars (section \ref{highmass}). In section \ref{helium}, I shall briefly address the problem of semi-convection in helium burning stars. Section \ref{discussion} will be devoted to a short discussion in terms of vibrational stability. My aim is here to emphasize the physical conditions for semi-convection to (possibly) develop, not to describe the modern ways of tackling the problem nor to present an exhaustive review of the theoretical works done since the sixties. These are presented and discussed in the next chapter \cite{zkm12} of which this constitutes a sort of preamble.

\section{A few basic points}\label{basic}

Let us recall here a few basic points affecting the stellar structure~:
\begin{enumerate}
\item Radiative temperature gradient

In a simplified way the radiative temperature gradient can be written
\begin{equation}
\nabla_{rad} \sim \frac{L}{m} \kappa
\end{equation}
where L is the luminosity and $\kappa$ the opacity.
This means that
\begin{itemize}
\item a large L/m value typical of nuclear burning cores is favorable to convection. The larger the temperature sensitivity of the nuclear energy production rate, the larger the L/m value;
\item a large opacity mostly found in ionization zones located in the outer layers leads to the presence of a convective envelope.
\end{itemize}
\item Temperature sensitivity of a nuclear reaction

The temperature sensitivity of a non resonant nuclear reaction involving the fusion of two nucleons $A_A$ and $A_a$ (see for instance \cite{clay68}) is given by
\begin{equation}
\nu = (\frac{d\lg{\varepsilon}}{d\lg{T}})_{\rho}
\end{equation}
where $\varepsilon$ is the nuclear energy production rate, T is the temperature and $\varrho$ the density. Its value strongly depends on the Gamow factor b
\begin{equation}
b = Z_A Z_a A_\mu^{1/2} \qquad {\rm with} \qquad  A_\mu = \frac{A_A A_a}{A_A + A_a}
\end{equation}
The $\nu$ value is then easily estimated from the relations
\begin{equation}
\nu = \frac{\tau - 2}{3} \qquad {\rm with} \qquad  \tau = \frac{3E_G}{kT} \qquad {\rm and} \qquad  E_G = (\frac{bkT}{2})^{2/3}
\end{equation}
where k is the Boltzmann constant and $E_G$ is the Gamow energy {\it i.e.} the {\it most effective energy} for the nuclear reaction to take place.
\\
\item Temperature sensitivity of the PP chain reactions

From the relations above it is evident that the smallest $\nu$ value will be obtained for the $(_1^1H,_1^1H)$ reaction. At a temperature of about $10^7K$ (kT $\sim$ 1 keV) $\nu_{11}$ is of the order of 4. As soon as the charge of the nucleons increases as in $(_2^3He,_2^3He)$, the Gamow factor b drastically increases and so does the temperature sensitivity. For kT equal to 1 keV, $\nu_{33}$ reaches a value of about 17.
\begin{itemize}

\item {\it PP chain operating out of equilibrium}

Near the end of the pre main sequence phase when the temperature at the center reaches a value of about $10^7K$ the hydrogen burning nuclear reactions start. In low mass stars hydrogen burning is largely dominated by the PP chain. The abundances of the nucleons involved in the PP chain, essentially that of $_2^3He$, are however still different from their equilibrium (or more precisely {\it stationary}) abundances. This means that $_2^3He$ is accumulating up to the point where its destruction rate will equal its formation rate. The temperature sensitivity is given by
\begin{equation}
\nu_{PP} = \nu_{11} \frac{\varepsilon_{11}}{\varepsilon} + \nu_{33} \frac{\varepsilon_{33}}{\varepsilon}
\end{equation}
Due to the high value of $\nu_{33}$ the resulting sensitivity is large and a convective core appears and remains during the whole process of reaching the equilibrium abundances.

\item {\it PP chain operating at equilibrium}

When $_1^2H$ and $_2^3He$ reach their equilibrium abundances {\it i.e.} when they are produced and destroyed at exactly the same rate, the whole PP chain is governed by the $(_1^1H,_1^1H)$ reaction (or {\it pp reaction}). That means that the $\nu$ value is low and the L/m ratio is not high enough to allow the presence of a convective core.

\item {\it PP chain operating with overshooting}

If a certain amount of overshooting, or any other extra mixing, is taken into account above the convective core boundary, the convective core present before ZAMS can be maintained during the whole main sequence phase. This is due to the fact that this extra mixing prevents $_2^3He$ from reaching its equilibrium abundance whatever the elapsed time since fresh $_2^3He$ is continuously brought into the core.

\end{itemize}

\item Temperature sensitivity of the CNO cycle reactions

Whatever the proton capture reaction involved in the CNO cycle, its sensitivity is high due to the large value of the Gamow factor. A convective core is present during the whole core hydrogen burning phase.
\\
\item Temperature sensitivity of the $3\alpha$ helium burning reaction

The $3\alpha$ reaction is a resonant reaction. Its temperature sensitivity (see \cite{clay68}) is given by
\begin{equation}
\nu_{3\alpha} = \frac{42.9}{T_8} - 3
\end{equation}
where $T_8$ is the temperature in $10^8K$. Such large values of $\nu$ involve the presence of a convective helium burning core in the whole stellar mass domain.
\\
\item Central temperature of MS stars

Assuming hydrostatic equilibrium, thermal equilibrium and radiative transfer, a dimensional reasoning easily leads to (see for instance \cite{kw96})
\begin{equation}
T \sim \frac{M}{R} \sim M^{1-\frac{\nu-1}{\nu+3}}
\end{equation}
Where M is the stellar mass and R is the surface radius. Whatever the way of burning hydrogen, through PP chain or CNO cycle, the exponent of M is positive which means that the central temperature increases with the stellar mass. This implies a progressive growth of the contribution of CNO in the nuclear reactions, from PP chain for low mass stars to CNO cycle in intermediate and massive stars. Another important effect is the increase of the contribution of the radiation pressure to the total pressure as the stellar mass increases.

\end{enumerate}

\section{Semi-convection in low mass stars}\label{lowmass}

\begin{figure}[H]
\begin{center}
\sidecaption[]
\includegraphics[angle=-90,width=1.0\textwidth]{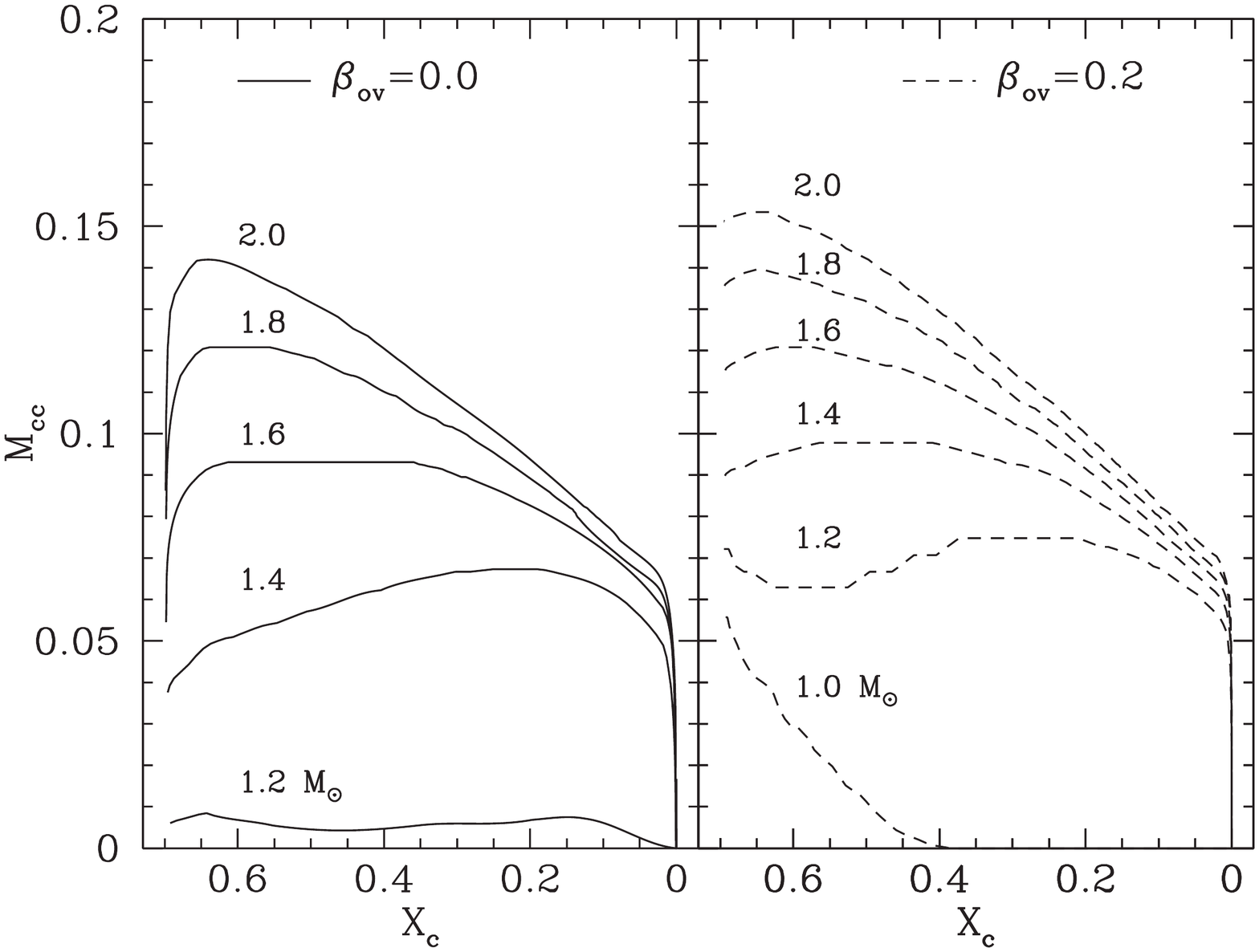}
\caption{Fractional convective core mass as a function of central hydrogen abundance $X_c$ for models in the mass range $1.0 M_{\odot} – 2.0 M_{\odot}$ computed without and with overshooting. The overshooting parameter $\beta$ (ratio of the overshooting distance and the minimum between the convective core radius and the local pressure scale height) is indicated in each panel. (From \cite{mig08})}
\label{Mig}
\end{center}
\end{figure}

In the mass range [$1.0M_{\odot}-2.0M_{\odot}$] the small convective core present while the PP chain elements are still reaching their equilibrium values, vanishes on the ZAMS. As the star evolves however the temperature at the center slowly increases which leads to a growing contribution of the CNO cycle to the nuclear reactions. The temperature sensitivity increases from a value of about 5 to a larger value typical of CNO reactions $(\sim15)$. A convective core appears and its mass extension increases as the CNO contribution increases. Figure \ref{Mig} shows the evolution of the fractional core mass extension with the central hydrogen abundance. This growing tendency is enhanced when models are computed with overshooting since a convective core is already present at the ZAMS due to the non equilibrium values of the $_2^3He$ abundance in the mixed region (see section \ref{basic}).

\begin{figure}[H]
\begin{center}
\includegraphics[width=5cm]{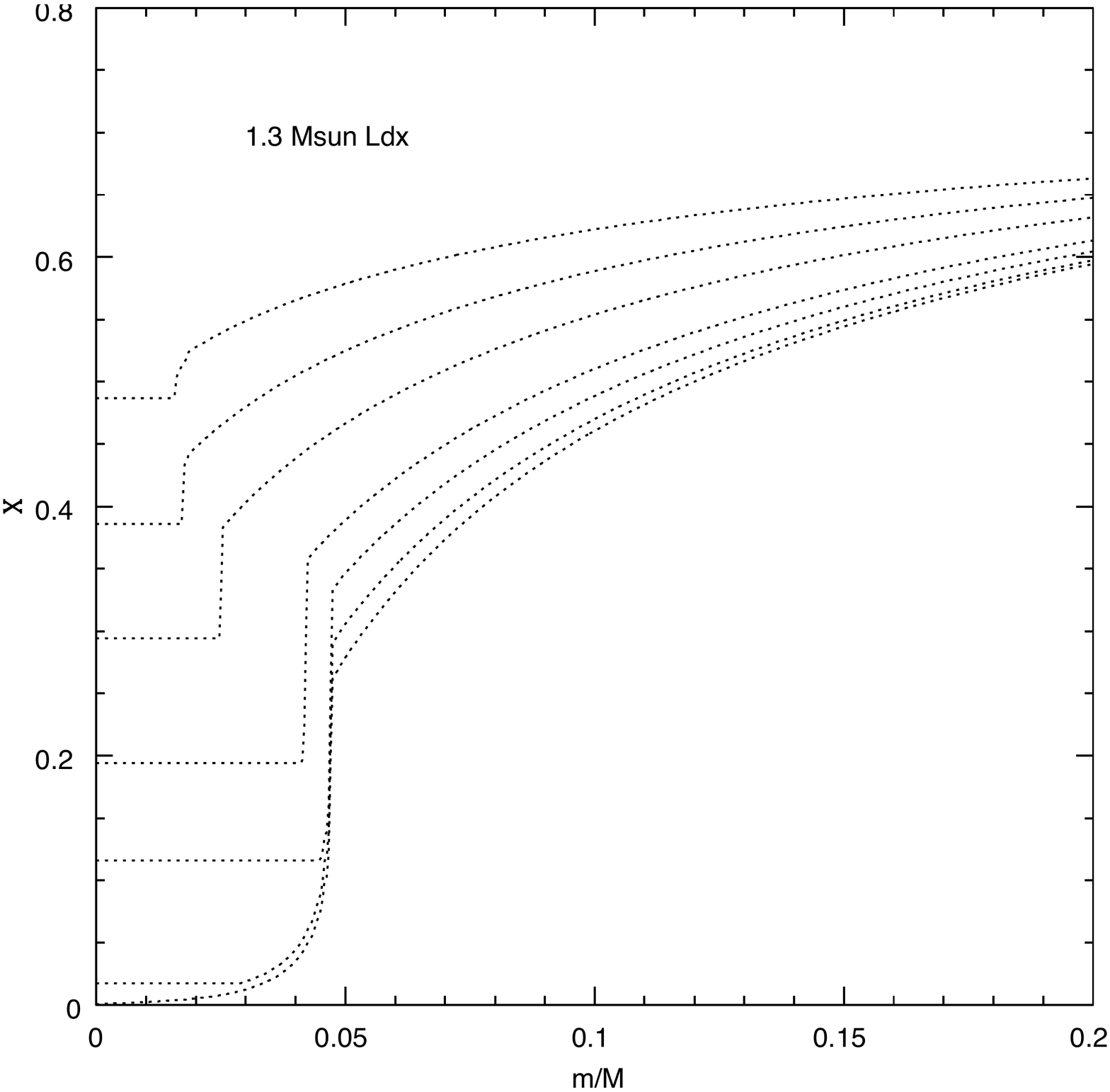}
\includegraphics[width=5cm]{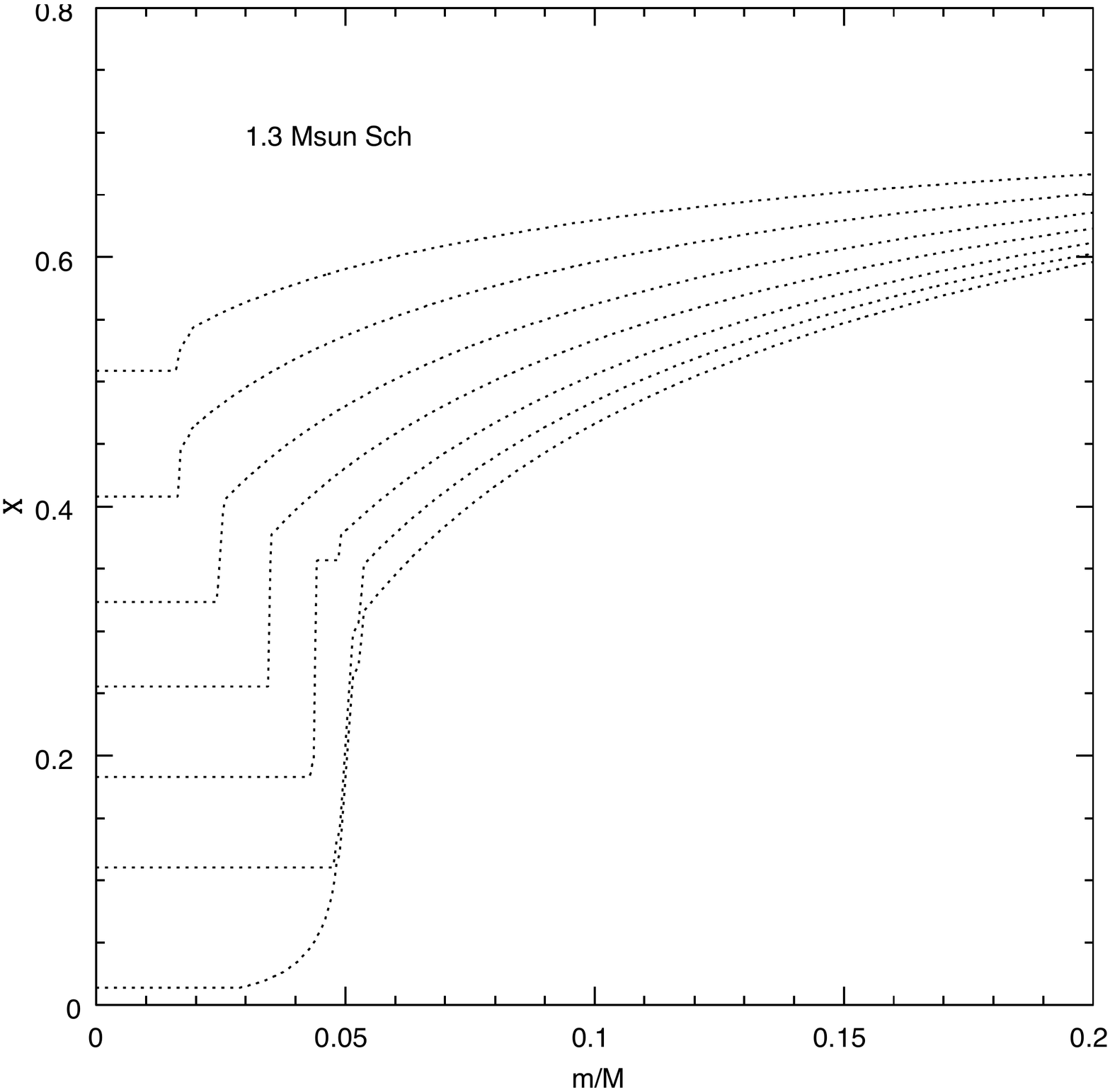}
\caption{Hydrogen profile in models of $1.3 M_{\odot}$ (X = 0.70, Z = 0.02) computed with the Ledoux criterion (left panel) and with the Schwarzschild criterion (right panel) in the fractional mass interval $[0,0.2]$.}
\label{X1.3}
\end{center}
\end{figure}

When models are computed with the Ledoux criterion such an evolution of the convective core leads to a well defined and numerically stable discontinuity at the convective core boundary as can be seen in the left panel in Fig. \ref{X1.3}. This is not the case when the Schwarzschild criterion is used (see the right panel in Figure \ref{X1.3}) since small convective shells form in the region of varying mean molecular weight ($\mu$-gradient region). The resulting effect is to create nearly neutral convective conditions in the whole $\mu$-gradient region and to reproduce more or less the Schwarzschild and H\"{a}rm solution of the problem of semi-convection (see section \ref{intro}). The number and the extent of these small convective shells are however dependent on the number of mesh points in the model in a somewhat erratic way. Although rather tempting this solution is dangerous especially because of the potential numerical diffusion induced by these moving convective layers. The mass domain of stars affected by this semi-convective problem is limited to [$1.0M_{\odot}-2.0M_{\odot}$] and the duration of the phase itself is only a small fraction of the main sequence lifetime since those intermediate convective zones appears mostly near the maximum extent of the convective core.

\begin{figure}[H]
\begin{center}
\sidecaption[]
\includegraphics[angle=-90,width=5.8cm]{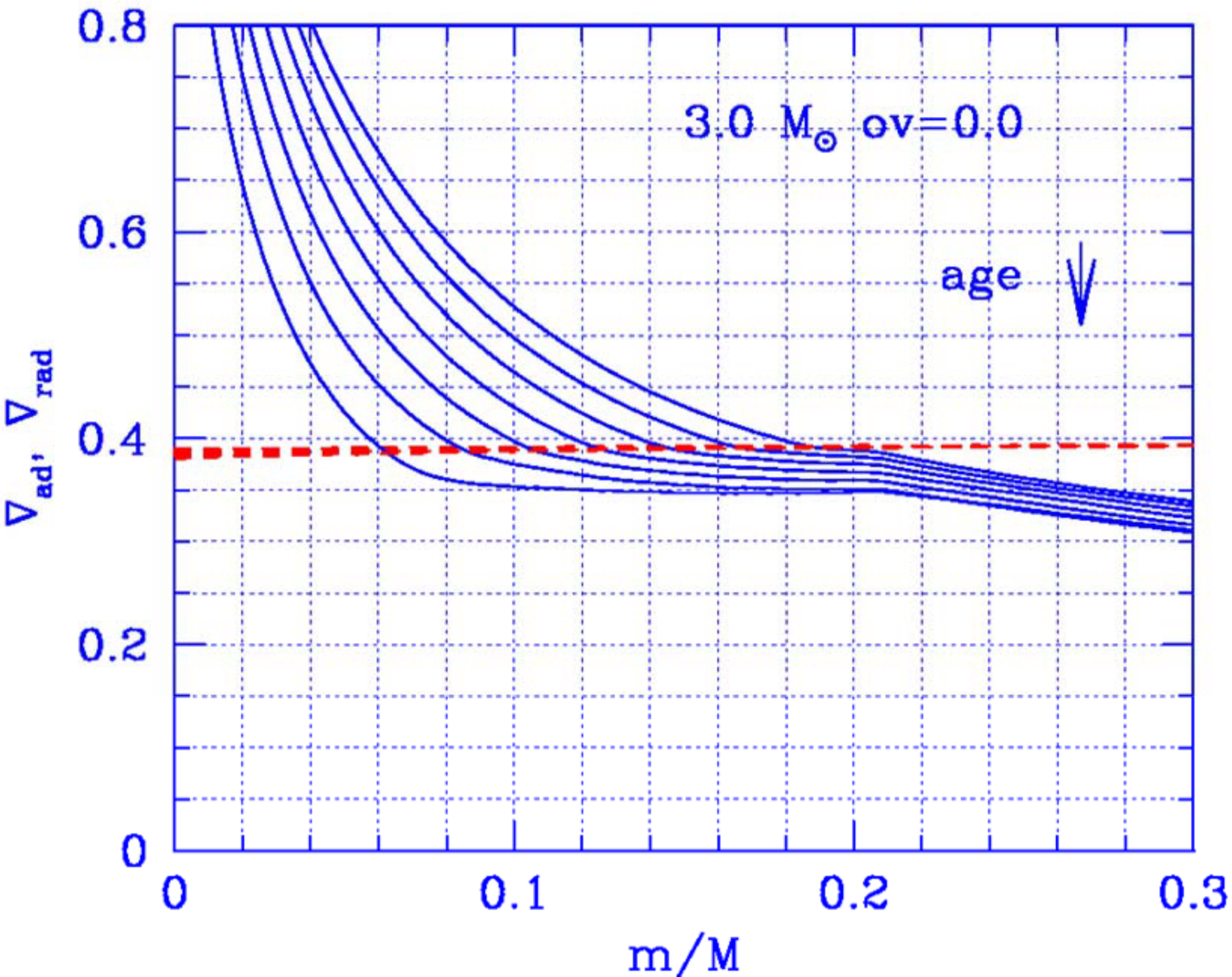}
\includegraphics[angle=-90,width=5.8cm]{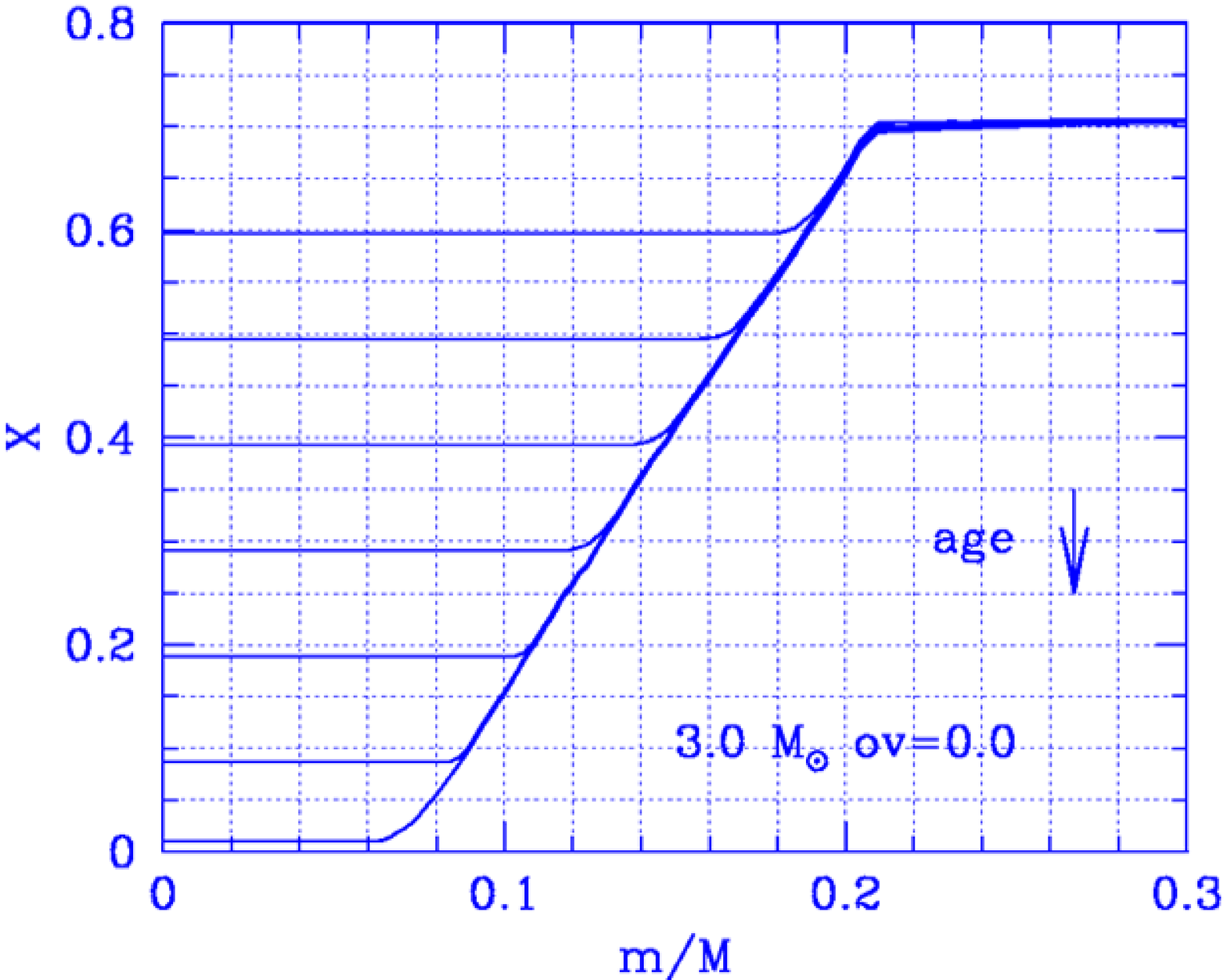}
\caption{Radiative and adiabatic temperature gradient distributions (left panel) and hydrogen profile (right panel) in models of $3M_{\odot}$ (X=0.70, Z=0.02) during MS. (From \cite{no04})}
\label{gradX}
\end{center}
\end{figure}

For intermediate mass stars more massive than $2.0M_{\odot}$ the problem of semi-convection disappears since hydrogen burning is dominated by the CNO cycle already at the ZAMS and as the evolution proceeds the temperature sensitivity does not change significantly. Since the hydrogen abundance X decreases in the convective core, $\nabla_{rad}$ decreases accordingly (see Figure \ref{gradX}) and the $\mu$-gradient region forms a smooth transition between the convective core and the homogeneous envelope.

\section{Semi-convection in massive stars}\label{highmass}

\begin{figure}[H]
\begin{center}
\sidecaption[]
\includegraphics[width=5cm]{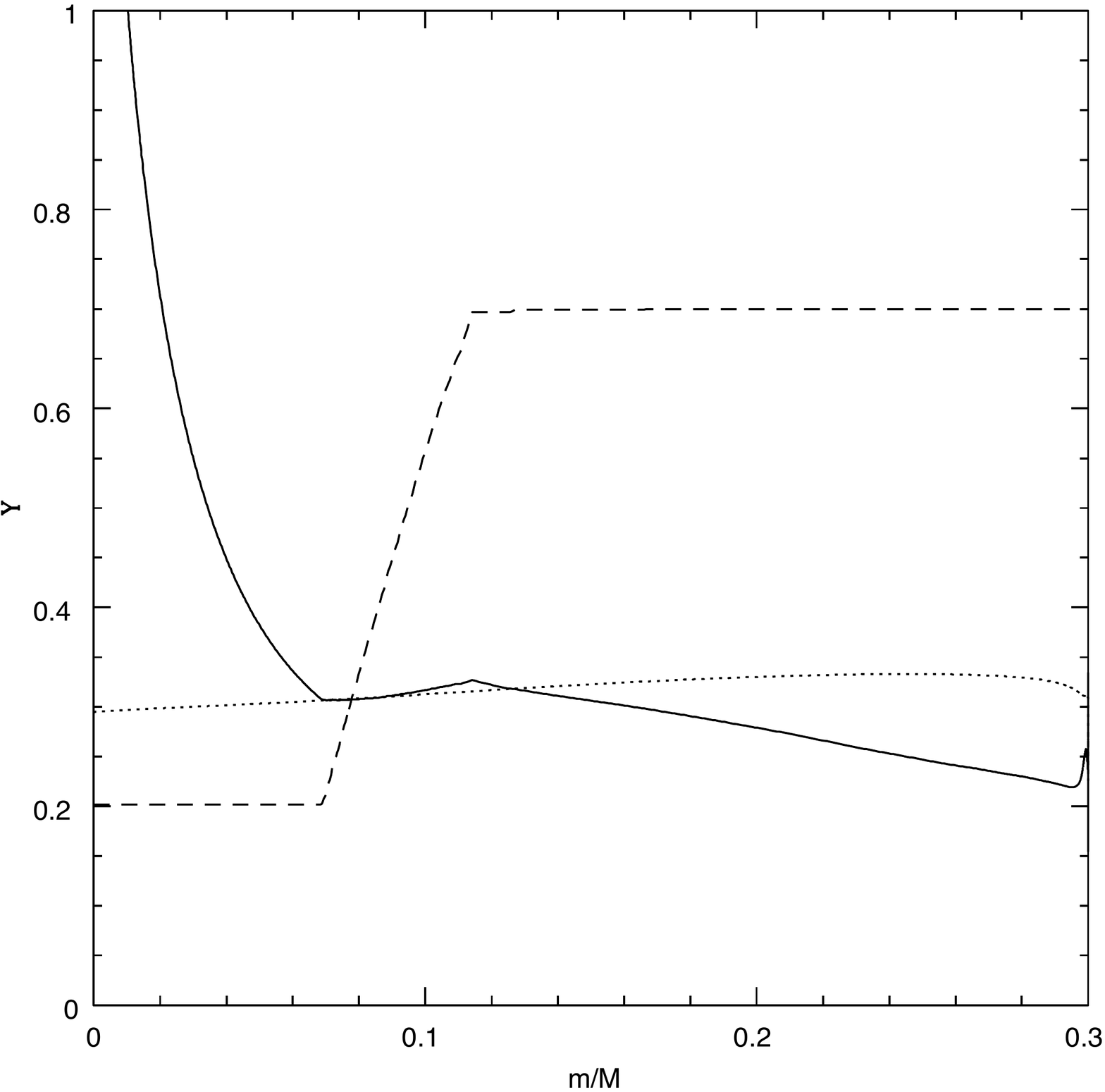}
\includegraphics[width=5cm]{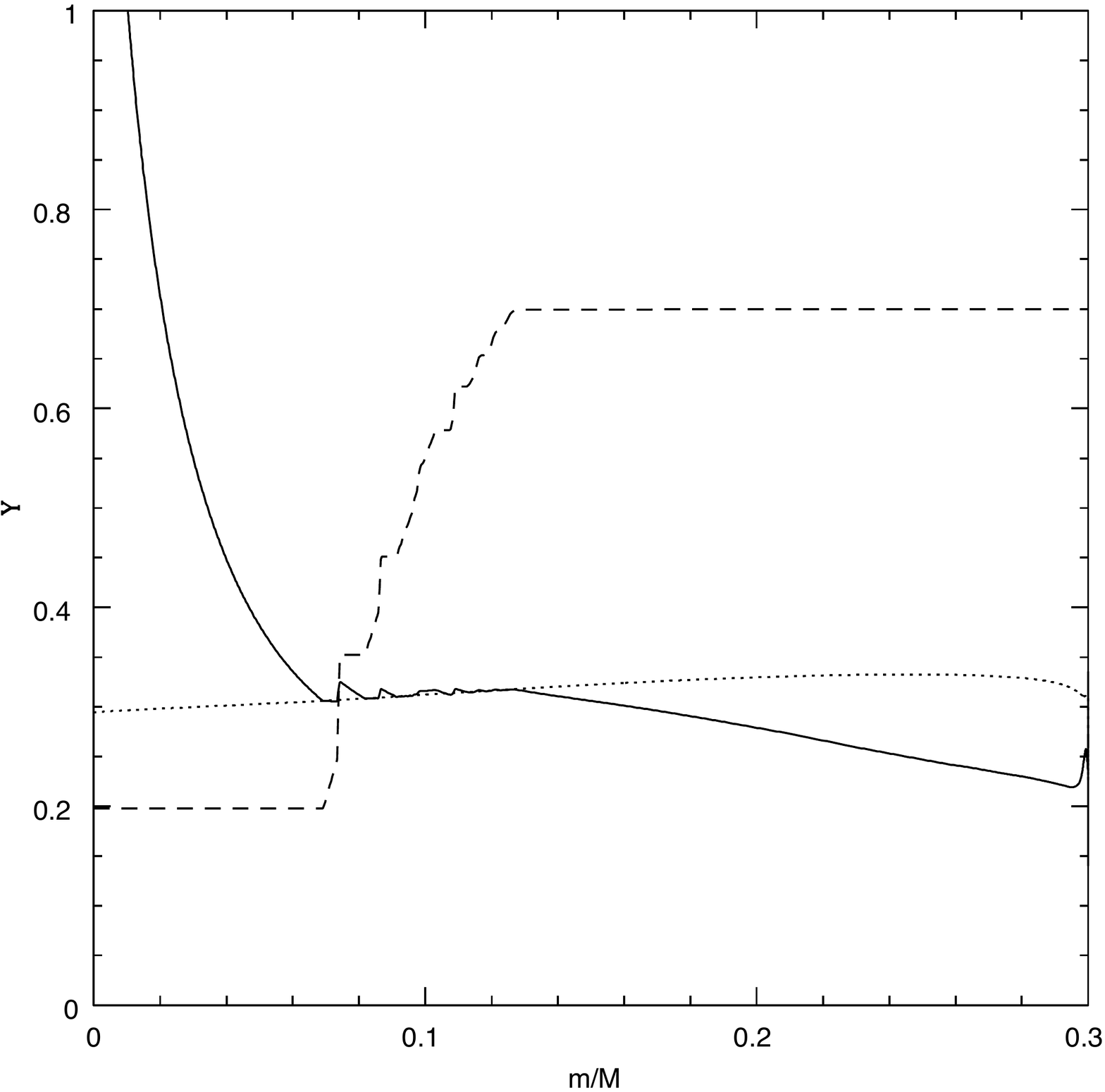}
\caption{Hydrogen profile (dashed line), $\nabla_{rad}$ (full line) and $\nabla_{ad}$ (dotted line) for an MS model with $X_c$=0.20) computed with the Ledoux criterion (left panel) and with the Schwarzschild criterion (right panel) in a sequence of $15M_{\odot}$ MS (X=0.70, Z=0.02).}
\label{Ldx-Sch15}
\end{center}
\end{figure}

Another problem arises for stellar masses larger than about $15M_{\odot}$. As recalled in section \ref{basic} the radiation pressure becomes a larger and larger contributor to the total pressure as the stellar mass increases. The resulting effect is a decrease of the adiabatic temperature gradient due to a progressive decrease of $\Gamma_2$ from 5/3 to 4/3 as $\beta$ (see section \ref{basic}) varies from 1 to 0. This effect favors convection and thus a larger mass extension of the convective core as M increases. As hydrogen burning proceeds the core mass extension decreases but $\nabla_{rad}$ and $\nabla_{ad}$ are extremely close to one another in the $\mu$-gradient region. With the Ledoux criterion the steepness of the $\mu$-gradient is such that the layers remain stable towards convection. On the contrary when adopting the Schwarzschild criterion small convective shells appear in $\mu$-gradient region. This is illustrated in Figure \ref{Ldx-Sch15} where the X profile, $\nabla_{rad}$ and $\nabla_{ad}$ are shown in the left and the right panels for a typical model ($X_c$=0.20) of a MS $15M_{\odot}$ computed with the Ledoux and with the Schwarzschild criterions respectively.

Near the end of core hydrogen burning a smell convective zone appears at the base of the homogeneous envelope in models computed with the Ledoux criterion. This does not affect the $\mu$-gradient profile which remains perfectly smooth with well defined junctions with the convective core and the homogeneous envelope plateaus during the whole MS. In models computed with the Schwarzschild criterion on the contrary the somewhat erratic behavior of transient convective shells leads to a step behavior of the X profile whose precise form depends on the adopted number of mesh points. The near equality of $\nabla_{rad}$ and $\nabla_{ad}$ in the $\mu$-gradient region tends to mimic the treatment of semi-convection advocated by \cite{sh58}.

As the mass increases [$M\geq 30M_{\odot}$] the problem becomes more and more crucial since the mass extension of the convective cores even grows with time during the MS. The resulting discontinuity in X at the convective core boundary leads to a situation somewhat similar to that discussed in section \ref{lowmass}.

\section{Semi-convection in helium burning stars}\label{helium}

\begin{figure}[H]
\begin{center}
\includegraphics[angle=180,width=5cm]{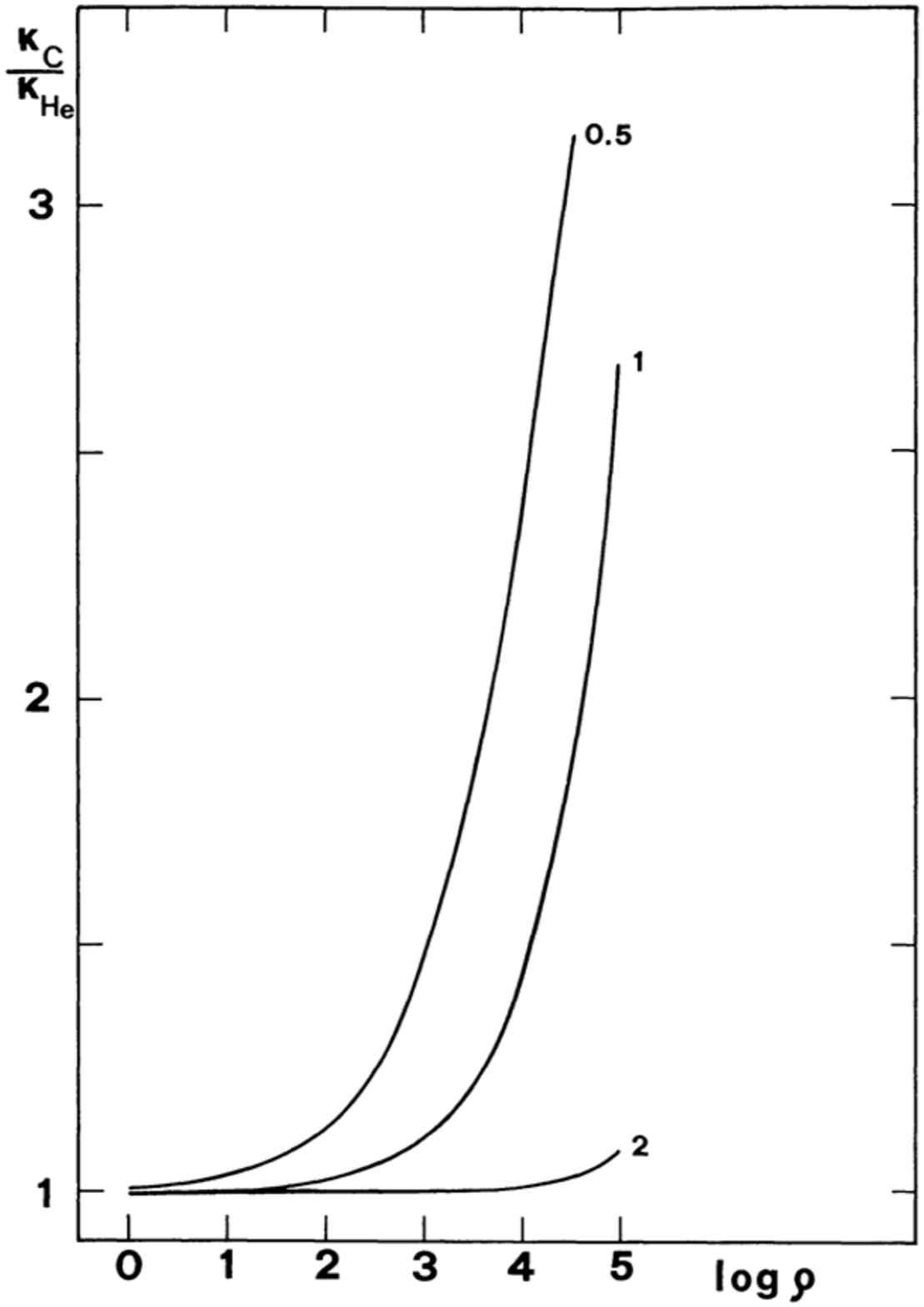}
\includegraphics[angle=180,width=5cm]{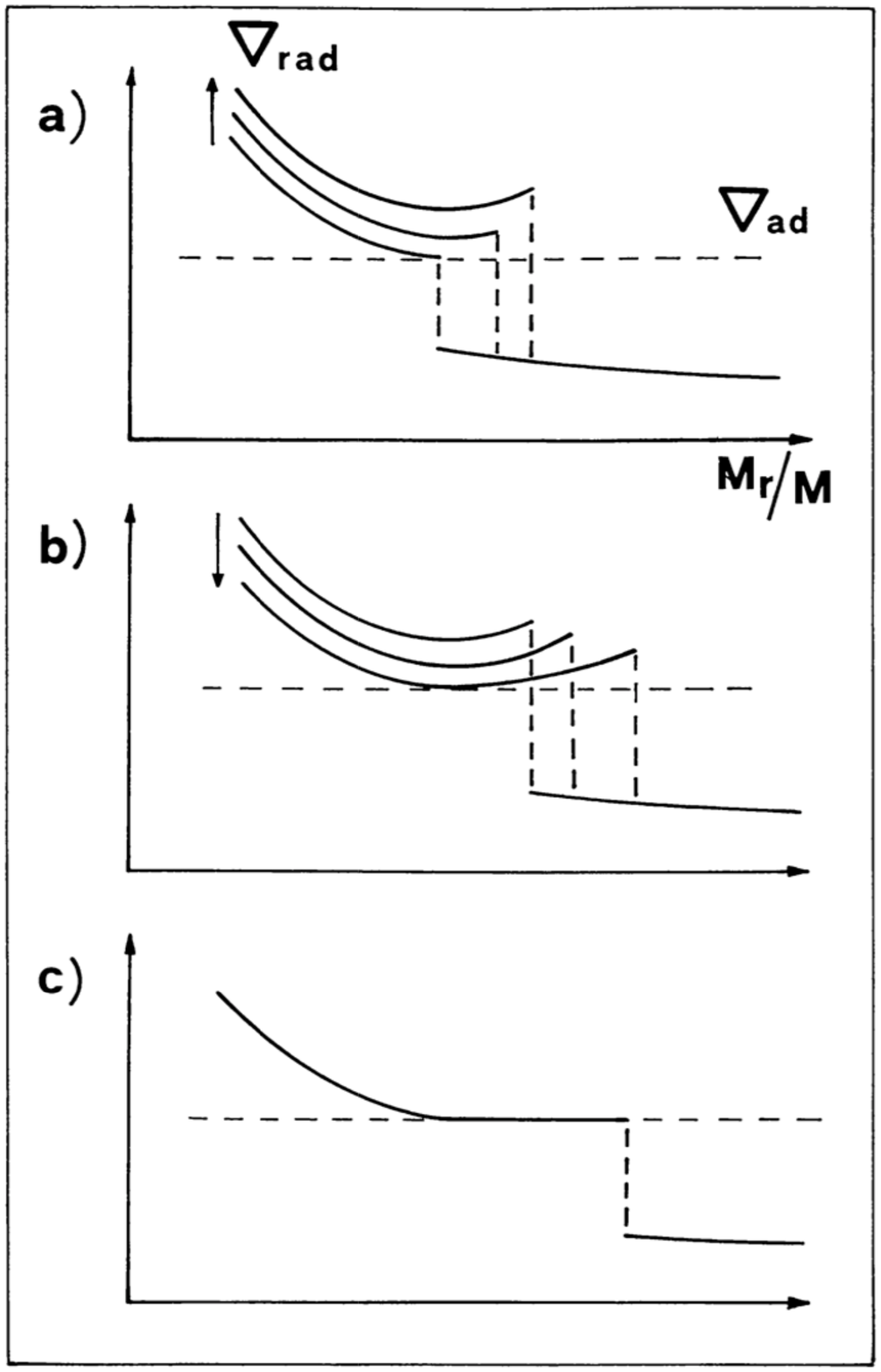}
\caption{Ratio of opacities in a pure carbon and a pure helium matter as a function of the logarithm of the density for labeled values of the temperature expressed in $10^8K$ (left panel; from \cite{cas171}). Mass distribution of $\nabla_{rad}$ and $\nabla_{ad}$ illustrating the problem of the minimum in $\nabla_{rad}$ during core helium burning. In the top and middle graphs $\nabla_{rad}$ increases and decreases with time respectively. The induced semi-convective mixing is shown in the bottom graph (right panel; from \cite{cas271}).}
\label{min}
\end{center}
\end{figure}

 In models of core helium burning massive stars the radiative opacity is dominated by electron scattering and is thus insensitive to the change in chemical composition produced by the transformation of helium into heavier $\alpha$-elements. This is not the case however for models burning helium at lower temperature and higher density as it is found in intermediate and low mass stars. The free-free transitions opacity depending on the charge of the ion is larger in carbon rich mixture than in helium richer ones. This is illustrated in the left panel of Figure \ref{min} taken from \cite{cas171}.

\begin{figure}[H]
\begin{center}
\sidecaption[]
\includegraphics[width=5cm]{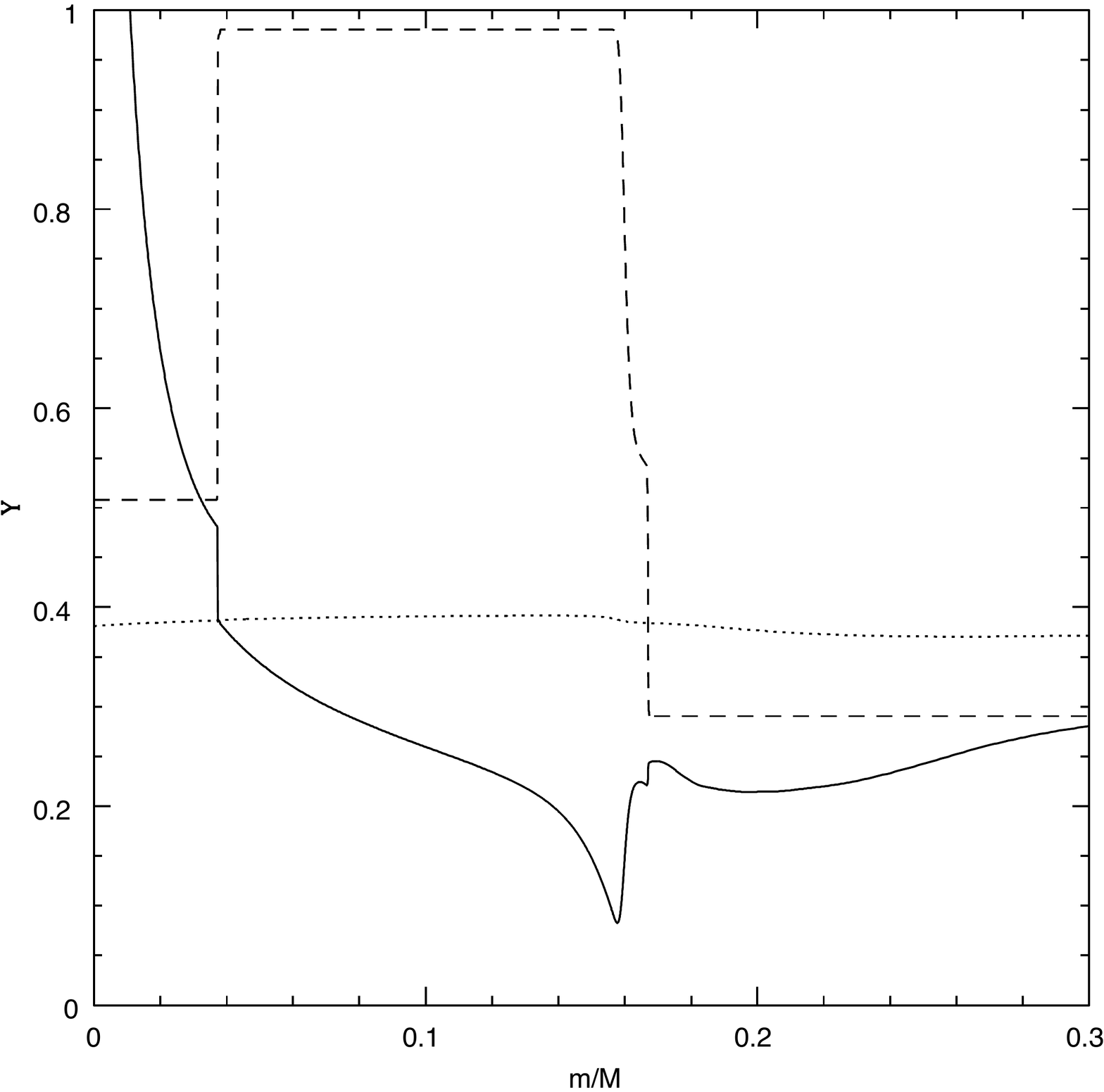}
\includegraphics[width=5cm]{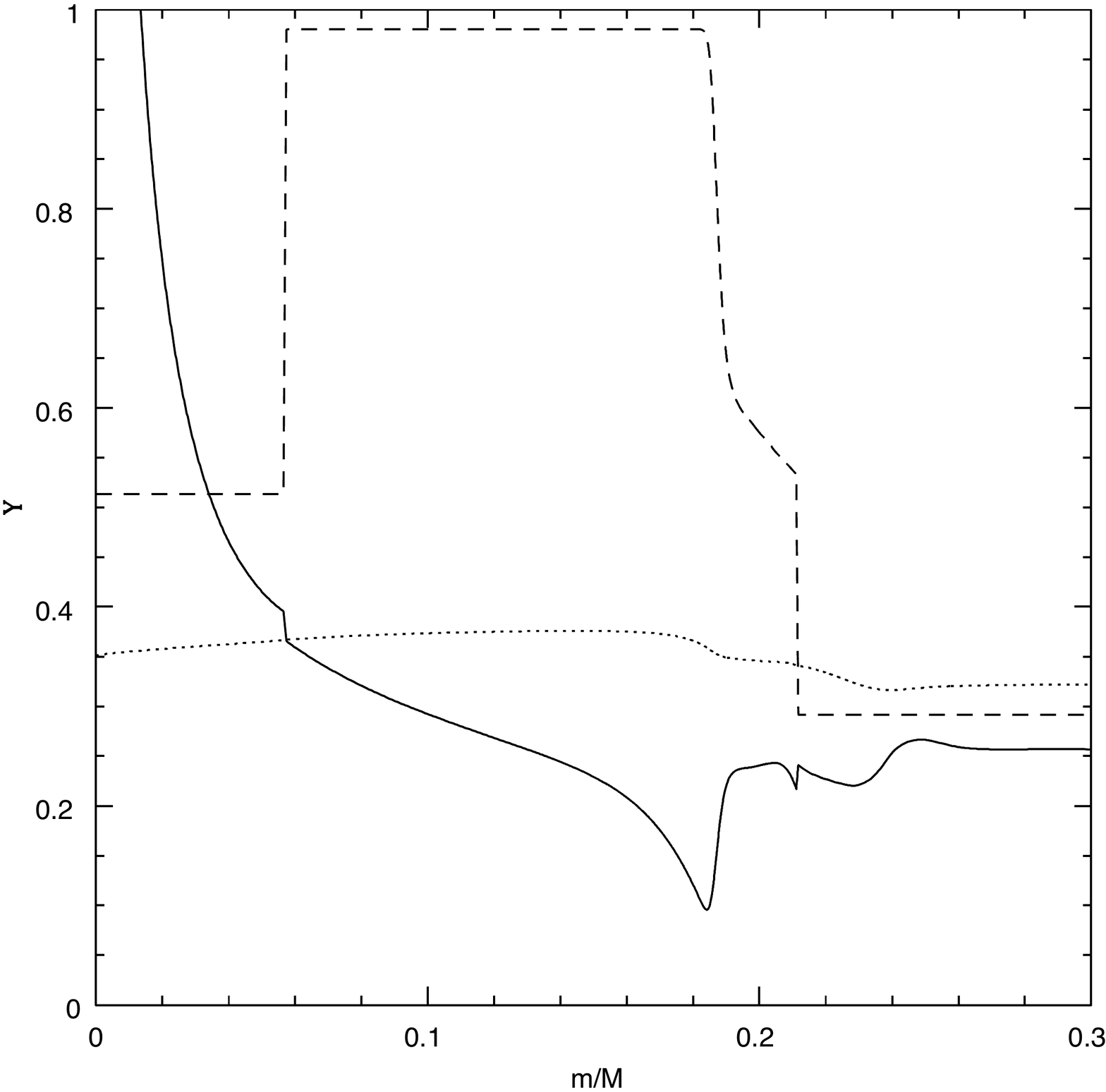}
\caption{Helium profile (dashed line), $\nabla_{rad}$ (full line) and $\nabla_{ad}$ (dotted line) as a function of the fractional mass [0,0.3] in models of $4M_{\odot}$ (left panel) and $8M_{\odot}$ (right panel) during core helium burning. A discontinuity in $\nabla_{rad}$ is visible at the boundary of the convective core.}
\label{He}
\end{center}
\end{figure}

Core helium burning stars all have a convective core as discussed in section \ref{basic}. As a result of the transformation of helium into carbon leading to a larger opacity the radiative temperature gradient increases in the convective core while it remains nearly constant just outside the convective boundary since the chemical composition is there unchanged. A discontinuity in $\nabla_{rad}$ tends to form at the core boundary as can be seen in Figure \ref{He} where $\nabla_{rad}$, $\nabla_{ad}$ and the helium profile as a function of the fractional mass [0,0.3] are given for a model of $4M_{odot}$ (left panel) and $8M_{\odot}$ (right panel) in the core helium burning phase.

As discussed in \cite{cas171} such a convective boundary is unstable since mixing a radiative layer close to the boundary with the homogeneous matter within the convective core enhances the carbon contain in the layer and makes it unstable towards convection. A sort of overshooting called by \cite{cas171} a {\it self driving mechanism} leads to the extension of the core up to a layer for which the Schwarzschild criterion is satisfied with the chemical composition of the convective core. This progressive outward shift of the core boundary meets however a new difficulty which was first analyzed in \cite{cas271}. The mass distribution of $\nabla_{rad}$ indeed presents a minimum before reaching the boundary. Wether $\nabla_{rad}$ increases or decreases with time the existence of this minimum prevents a coherency in the determination of the convective boundary as can be seen in the top and middle graphs in Figure \ref{min} (right panel). The bottom graph shows the resulting {\it induced semi-convection} as proposed in \cite{cas271}.

\section{Discussion in terms of vibrational stability}\label{discussion}
I very briefly discuss here some earlier works on the stability aspects relevant to semi-convection. Classical convection is due to a dynamical instability of $g^-$ modes while overstable convection present in layers of varying mean molecular weight results from a vibrational instability of dynamically stable $g^+$ modes trapped in the $\mu$-gradient region \cite{ka66}. The physical structures discussed in the above sections are generally favorable to the existence of such trapped $g^+$ modes. It is thus tempting to interpret a semi-convective mixing as a mixing resulting from the instability of these $g^+$ modes. The precise nature of the mixing is however far from straightforward. Since the destabilizing term tends to vanish when $\nabla_{rad}$ and $\nabla_{ad}$ are equal, it would be safe to say that a mixing tending to this equality would lead to a stable situation. Another mechanism acting in massive stars was proposed by \cite{ga70} who showed that transient convective shells would progressively move across and partially mix the $\mu$-gradient region until the Schwarzschild neutrality condition is met.

As already noted in section \ref{intro} attempts have however been made by numerous stellar evolution scientists at forming semi-mixed regions either neutral towards the Schwarzschild or towards the Ledoux criterion. As pointed by Ledoux \cite{led47} adopting the Schwarzschild criterion as a neutrality condition in a semi-convective region has the advantage of leading to a dynamically stable situation while the Ledoux criterion leads to marginally dynamically unstable conditions which would inevitably lead to a full convective mixing. {\it Ledoux was indeed the first to advocate in this context the use of the Schwarzschild criterion and NOT the Ledoux criterion as a neutrality condition in semi-convective zones}.

Even trapped modes must be checked through a full stability analysis. In the case of massive MS stars \cite{gn76} showed that low order $g^+$ modes of high spherical harmonic degree could be trapped in the $\mu$-gradient region. The timescales were found to be of the order of $10^3$ to $10^4$ yr which made them good candidates for a partial semi-convective mixing.

For low mass MS stars a similar analysis was performed \cite{gn77} and unstable trapped low order $g^+$ modes of high degree were also found. A very narrow mass range near $1.1M_{\odot}$ was however affected by this instability. Moreover unstable modes were only found during a short part of the main sequence phase near the maximum extent of the convective core.

Although impossible to draw a full picture of semi-convection from these stability analyses it is interesting to point out that they seem to converge towards a partial chemical mixing of the semi-convective layers ensuring their being neutral with respect to the Schwarzschild criterion.


\begin{thebibliography}{}

\bibitem{cas171} Castellani, V., Giannone, P., Renzini, A.: Overshooting of convective cores in helium burning horizontal branch stars. Astrophys. and Space Sc. {\bf 10} 340-349 (1971)

\bibitem{cas271} Castellani, V., Giannone, P., Renzini, A.: Induced semi-convection in helium burning horizontal branch stars. Astrophys. and Space Sc. {\bf 10} 355-362 (1971)

\bibitem{clay68} Clayton, D. D.: Principles of stellar evolution and nuclear synthesis. McGraw-Hill Book Company, Chap. 4 (1968)

\bibitem{ga70} Gabriel, M.: On the mechanism of formation of semi-convective zones in stars. Astron. Astrophys. {\bf6} 124-129 (1970)

\bibitem{gn76} Gabriel, M. Noels, A.: Stability of a $30M_{\odot}$ star towards $g^+$ modes of high spherical values. Astron. Astrophys. {\bf53} 149-157 (1976)

\bibitem{gn77} Gabriel, M. Noels, A.: Semi-convection in stars of about $1M_{\odot}$. Astron. Astrophys. {\bf54} 631-634 (1977)

\bibitem{ka66} Kato, S.: Overstable stability in a medium stratified in mean molecumar weight. Pub. Astron. Soc. Japan {\bf 18} 374-383 (1966)

\bibitem{kw96} Kippenhahn, R., Weigert, A.: Stellar structure and evolution. Springer (1996)

\bibitem{led47} Ledoux, P.: Stellar models with convection and with discontinuity of the mean molecular weight. Astrophys. Jour. {\bf 105} 305-321 (1947)

\bibitem{mig08} Miglio, A., Montalb\`{a}n, J., Noels, A., Eggenberger, P.: Probing the properties of the convective cores through g modes: high-order g modes in SPB and $\gamma$ Doradus stars. Mon. Not. R. Astron. Soc. {\bf 386} 1487-1502 (2008)

\bibitem{no04} Noels, A., Montalb\`{a}n, J., Maceroni, C.: A-type stars:bevolution, rotation and binarity. In: Zverko, J., Ziznovsky, J., Adelman, S.J., Weiss, W.W. (eds) The A-star puzzle, IAU Symp. {\bf 224} 47-57 (2004)

\bibitem{sh61} Sakashita, S., Hayashi, C.: Internal structure of very massive stars. Prog. of Theor. Phys. {\bf 26} 942-946 (1961)

\bibitem{schw58} Schwarzschild, M.:  {\em Structure and evolution of the stars}, Dover Publ. Inc. New York (1958)

\bibitem{sh58} Schwarzschild, M., H\"{a}rm, R.: Evolution of Very Massive Stars. Astrophys. Jour. {\bf 128} 348-360 (1958)

\bibitem{zkm12} Zaussinger. F., Kupka. F., Muthsam, H.J.: Semi-convection. This volume (2012)

\end{thebibliography}
\end{document}